\documentclass[a4paper]{article}
\usepackage{spconf}
\usepackage{textcomp}
\usepackage{soul}
\usepackage{graphicx}
\usepackage{amsmath,amssymb}
\usepackage{amsthm}
\usepackage[top=25mm,bottom=25mm,left=22mm,right=22mm]{geometry}
\usepackage[english]{babel}
\usepackage[lofdepth,lotdepth]{subfig}
\usepackage{color}
\usepackage[labelfont=bf]{caption}
\usepackage{blindtext}

\usepackage[ruled,vlined]{algorithm2e}
\usepackage{todonotes}

\newcommand{\cel}{{\sf c}}

\newcommand{\bs}{\boldsymbol}
\newcommand{\bb}{\mathbb}
\newcommand{\cl}{\mathcal}

\newcommand{\ts}{\textstyle}
\newcommand{\ie}{\emph{i.e.}, }

\newcommand{\tiid}{%
    \ifmmode
        \mathrm{iid}%
    \else%
        iid\xspace%
    \fi%
}

\newcommand{\tm}{\times}

\newcommand{\jj}{\mathrm{j}}

\newcommand{\ong}{\emph{on-the-grid} }
\newcommand{\ofg}{\emph{off-the-grid} }

\newcommand{\sps}[3]{#3\langle#1,\,#2 #3\rangle}
\newcommand{\amax}[1]{\underset{\hspace{1mm}#1\hspace{1mm}}{\text{arg\,max}\hspace{1mm}}}
\newcommand{\amin}[1]{\underset{\hspace{1mm}#1\hspace{1mm}}{\text{arg\,min}\hspace{1mm}}}

\newcommand{\bpsi}{\bs \psi}

\newcommand{\bphi}{\bs \phi}

\newcommand{\OP}{\Omega_{\cl P}}
\newcommand{\OR}{\Omega_{\cl R}}
\newcommand{\OV}{\Omega_{\cl V}}
\newcommand{\omp}{\bs \omega}
\newcommand{\omr}{\bar r}
\newcommand{\omv}{\bar v}

\newcommand{\rr}{\mathrm{r}}
\newcommand{\rv}{\mathrm{v}}

\definecolor{darkgreen}{rgb}{0.1,0.6,0.2}

\title{Sparse Factorization-based Detection of Off-the-Grid\\ 
Moving targets using FMCW radars\vspace{-3mm}}
\name{Gilles Monnoyer de Galland$^\star$, Thomas Feuillen$^\star$, Luc Vandendorpe$^\star$, Laurent
Jacques$^\dagger$\thanks{GM and LJ are funded by the Belgian FNRS.}\vspace{-3.5mm}}
\address{\ninept $^\star$ELEN. $^\dagger$INMA.  ICTEAM, UCLouvain, Belgium\vspace{-3mm}}

\AtBeginDocument{%
   \setlength\abovedisplayskip{100pt}
   \setlength\belowdisplayskip{100pt}}
\begin{document}
\ninept
\maketitle



\begin{abstract}

In this paper, we investigate the application of continuous sparse signal reconstruction algorithms for the estimation of the ranges and speeds of multiple moving targets using an FMCW radar.
Conventionally, to be reconstructed, continuous sparse signals are approximated by a discrete representation. 
This discretization of the signal's parameter domain leads to mismatches with the actual signal.
While increasing the grid density mitigates these errors, it dramatically increases the algorithmic complexity of the reconstruction. 
To overcome this issue, we propose a fast greedy algorithm for \ofg detection of multiple moving targets. 
This algorithm extends existing continuous greedy algorithms to the framework of factorized sparse representations of the signals. 
This factorized representation is obtained from simplifications of the radar signal model which, up to a model mismatch, strongly reduces the dimensionality of the problem. 
Monte-Carlo simulations of a $K$-band radar system validate the ability of our method to produce more accurate estimations with less computation time than the \ong methods and than methods based on non-factorized representations. 
\end{abstract}


\section{Introduction}
\label{sec_intro}
Sparse signal processing grew in interest for radar applications as it enables using Compressive Sensing (CS)
to reduce the amount of acquired data required to detect the targets~\cite{baraniuk2007, ahmad2012, feuillen2018}, or to achieve super-resolution~\cite{herman2009, zheng2017}. 
In this paper, we consider the estimation of the ranges and the radial speeds of several targets with a Frequency Modulated Continuous Wave (FMCW) mono-static radar emitting linear chirp modulated waveforms.
The sampled measurement signal from this system is assumed to be a linear combination of a few waveforms, each corresponding to the presence of one of the targets.
These waveforms (or atoms) are taken from a continuous parametric dictionary, in which the measurement is represented.
The detection of targets is thereby formulated as the reconstruction of a continuous sparse signal in the range-Doppler domain which models the observed scene.
Traditional approaches for sparse signal reconstruction capitalize on a discretization of the parameter domain and on the assumption that all the target parameters match the resulting grid~\cite{candes2008}.
Yet, reconstructions from such discretized models are affected by \emph{grid errors} that limit both the precision and the resolution of the algorithms~\cite{azodi2016, gribonval2019}. 
Although a denser grid reduces this effect, it tremendously increases the dimensionality of the problem to solve. 
This recently motivated many contributions on continuous sparse signal reconstruction algorithms, notably used to perform \ofg estimations.
For instance, in~\cite{tang2013, mishra2014, mishra2015}, the authors reformulate the reconstruction problem as a Semi-Definite Program.
In~\cite{keviren2017, traonmilin2018, traonmilin2019}, an approximated formulation of the sparse reconstruction is solved by a gradient descent.
Nonetheless, the high computational complexity of these methods makes them not suitable for real-time applications. 

In~\cite{simoncelli2011}, a continuous version of the Basis Pursuit~\cite{BP} is derived from the definition of an interpolated model that approximates the \ofg atoms. 
In~\cite{knudson2014, duarte2014}, the authors similarly designed a continuous adaptation of the Orthogonal Matching Pursuit (OMP)~\cite{OMP}, namely the Continuous OMP (COMP), using the same interpolation concept. 
The low complexity of this algorithm makes it of high interest for real-time radar applications. 
Still, COMP has only been formulated for the estimation of scalar parameters.
In this paper, we propose a formulation of COMP for the estimation of both the ranges and speeds of targets from the received signal of an FMCW radar.
%
Conventionally, such radar signals are modelled by applying known simplifications that enable a factorized expression of the radar atoms~\cite{winkler2007, lutz2014}.
The resulting atoms factorize in a product of two sub-atoms, respectively depending on the ranges and on the speeds of the targets. 
Among \ong algorithms, the 2D factorized OMP (F-OMP) presented in~\cite{fang2015} can take advantage of the factorization to reduce the dimensionality of problems. 
The simplification of the radar atoms, however, causes \emph{model mismatch} which generates distortions in the reconstruction process~\cite{liu2010, bao2014, feuillen2016}.
Thereby, the conventional application of F-OMP to radar systems is affected by both the \emph{grid errors} and the \emph{model mismatch}.

Our main contribution is to combine the concepts of interpolation and factorization to build the Factorized COMP (F-COMP) that enables a fast and accurate estimation of target parameters with FMCW radars.
Our simulations show the superiority of using low-density grids with interpolated models --- and hence, \ofg algorithms --- instead of denser grids with \ong algorithms. 
This superiority still holds when the reconstruction is affected by the model mismatch appearing from the factorization of the radar model. 
From the simulations, we also evaluate the possible degradation of the performance of F-COMP due to the model mismatch and its sensibility to radar system parameters.

\vspace{-1.5mm}
\paragraph*{Notations:} Matrices and vectors are denoted by bold symbols, $\jj=\sqrt{-1}$, and $\cel$ is the speed of light. The scalar product between the vectors $\bs{a}$ and $\bs{b}$ reads $\sps{\bs{a}}{\bs{b}}{}$. The transpose and conjugate transpose of a matrix $\bs{A}$ are $\bs{A}^\top$ and $\bs{A}^H$, respectively. The modulo operator is ${\rm mod}$, $\circledast$ is the convolution operator, $\odot$ is the Hadamard (element-wise) product, $\|\cdot\|_F$ is the Frobenius norm, $[N] := \{1,\cdots,N\}$, and $\bb C \cl N(0,\sigma^2)$ is the centered complex normal distribution of variance $\sigma^2$. 



\section{Signal Model and Approximations}
\label{sec_sig}
In this section, we formulate the radar signal as a linear combination of atoms parameterized by continuous parameters.
Next, we apply simplifications to the resulting model to build a factorized expression of these atoms.
We finally sample the parameter domain to obtain a grid from which we specify interpolation-based approximations for both the exact and factorized expressions of the atoms.
\subsection{Radar signal model:}
The transmitting antenna of an FMCW radar continuously emits a modulated wave which can be expressed by~\cite{hymans1960}
\begin{equation}
    \ts s_T(t) = \exp{\big(\jj2\pi\ts\int_0^t f_c(t')\mathrm{d}t'\big)},
    \label{eq_sT}
\end{equation}
where the transmission power is arbitrarily set to 1, and $f_c(t)$ is the instantaneous carrier frequency at instant $t$. For a linear chirp modulation, we have
\begin{equation}
    \ts f_c(t) = f_0 + B (\frac{t}{T_c}\!\!\!\!\mod 1),
    \label{eq_fc}
\end{equation}
where $f_0$ is the lowest frequency, $B$ is the bandwidth of the transmitted signal and $T_c$ is the chirp duration.

We consider $K$ point targets located at ranges $\{r_k\}_1^K \subset \cl R\subset \bb R^+$ and moving with radial speeds $\{v_k\}_1^K \subset \cl V\subset \bb R$.
The speed are assumed to be constant during an acquisition frame.
The target parameters are thus known to lie within the parameter domain $\cl P := \cl X \times \cl V \subset \bb R^2$.
The signal at the receiver side is expressed by
\begin{equation}
    \ts s_R(t) = s_T(t) \circledast\big[\sum_{k=1}^K\alpha_k\delta\big(t-\tau_k(t)\big)\big],
    \label{eq_channel}
\end{equation}
where 
$\alpha_k$ are scattering coefficients modelling all effects occurring in the wave reflection process, including the radar cross sections. 
We consider an ideal noiseless and clutterless scenario. 
$\tau_k (t)$ is a \textit{Delay-Doppler} term defined by 
\begin{equation}
\ts \tau_k(t) = \frac{2}{\cel}(r_k + v_k t).
\label{eq_taudef}    
\end{equation}
After coherent demodulation, the received baseband signal is
\begin{equation}
    \ts e_R(t) = \sum_{k=1}^K\alpha_k \exp\big(\ts -\jj 2\pi \big(f_c(t)\tau_k(t) - \frac{B}{2T_c}(\tau_k(t))^2\big)\big).
    \label{eq_baseband}
\end{equation}
The above signal is sampled at rate $1/T_s$ with $M_s$ samples acquired per chirp and $M_c$ chirps in total. 
There are $M:= M_sM_c$ acquired samples in total.
The resulting sampled measurement vector $\bs y \in \bb C^{M_cM_s}$, such that $y_{m_cM_s+m_s}=e_R(m_cT_c+m_sT_s)$, reads
\begin{equation}
    \ts \bs y = \sum_{k=1}^K\alpha_k\bs a(r_k, v_k),
    \label{eq_complete_y}
\end{equation}
where, for all $(r, v) \in \cl P$, the vector $\bs a(r, v)$ is an atom of the continuous radar sensing dictionary, defined by $\cl D := \{ \bs a(r, v) : (r,v)\in\cl P \}$. 
Thus, \eqref{eq_complete_y} formulates the complete radar signal as a linear combination of atoms taken in the parametric dictionary $\cl D$.

\subsection{Model factorization} 
To reduce the dimensionality of the problem solved in Sec. \ref{sec_alg}, we aim to decouple the range $r$ and the velocity $v$ in the atom $\bs a(r, v)$. 
From the sampling of \eqref{eq_baseband}, we can decompose $\bs a(r, v)$ as 
\begin{equation}
    a_{m_cM_s+m_s}(r, v) = \psi_{m_s}(r, v)\phi_{m_c}(v)\theta_{m_s,m_c}(r, v), 
\end{equation}
and
\begin{align}
    \ts \psi_{m_s}(r, v) = & \ts \exp\big(-\jj2\pi\frac{B}{M_s}\frac{2(r + \gamma v)}{\cel} m_s \big), \label{eq_psidef} \\
    \ts \phi_{m_c}(v) = & \ts \exp\big(-\jj2\pi f_0T_c \frac{2 v}{\cel} m_c\big) , \label{eq_phidef}\\
    \ts \theta_{m_s,m_c}(r, v) = & \ts \exp\big(-\jj\frac{2\pi}{\cel}\frac{B}{M_s}\big(\frac{r}{\cel T_s}+m_s\big)v(m_cT_c+m_sT_s)\big)
    \cdot \ts \exp\big(\jj\pi\frac{B}{M_sT_s}\frac{v^2}{\cel^2}(m_cT_c+m_sT_s)^2\big),  \label{eq_thetadef}
\end{align}
with $\gamma = \frac{f_0M_sT_s}{B}$.
Note that the above equations remain identical when time gaps are inserted between chirps (\ie $T_c>M_sT_s$).

Typically, factorizing the expression in \eqref{eq_complete_y} relies on a few assumptions~\cite{liu2010, bao2014, feuillen2016}: (i) $\cl R$ is such that for all $r\in\cl R$, $r<\cel T_s$, (ii) $\cl V$ is such that for all $v\in\cl V$, $|v| \leq \frac{\cel}{4f_0T_c}$ and (iii), the transmitted signal is narrowband and such that $M_c\frac{B}{f_0}<1$. 
The conditions above enable the approximation $\theta_{m_cM_s+m_s}(r, v) \simeq 1$. 
Therefore, given $\bs Y\in\bb C^{M_s\times M_c}$ the matrix-shaped $\bs y$ with $Y_{m_s, m_c} = y_{m_cM_s+m_s}$, the factorized model reads 
\begin{equation}
    \ts \bs Y \simeq \sum_{k=1}^K \alpha_k \bs A(r_k, v_k),
    \label{eq_Yfact}
\end{equation}
Defining the sub-atoms $\bs \psi(r+\gamma v) := \big(\psi_1(r,v), \cdots, \psi_{M_s}(r, v)\big)^\top$ and $\bs \phi(v) := \big(\phi_1(v),  \cdots ,  \phi_{M_c}(v)\big)^\top$, we have
\begin{equation}
    \bs A(r, v) := \bpsi(r+\gamma v)\bphi(v)^\top = \bpsi(r')\bphi(v)^\top.
    \label{eq_fact_atom}
\end{equation}
In \eqref{eq_fact_atom}, $\bphi(v)$ links $\cl V$ to the rows of $\bs A(r, v)$ and $\bpsi(r + \gamma v)$ links $\cl R':=\{r' = r+\gamma v: r\in\cl R, v\in\cl V\}$, to the columns of $\bs A(r, v)$. Hence \eqref{eq_Yfact} approximates the radar signal as a linear combination of atoms $\bs A(r_k, v_k)$ that are factorized by \eqref{eq_fact_atom} and such that $r'$ and $v$ are decoupled. Note that $(r, v)$ is retrieved from $(r', v)$ by substracting the deterministic offset $\gamma v$ from $r'$.



\subsection{Interpolated sparse representation:} 
Conventional methods to recover parameter values from signals modelled as \eqref{eq_complete_y} require the assumption that these parameters are taken from a grid which results from the sampling of $\cl P$.
The greedy algorithm ``Continuous OMP" (COMP)~\cite{knudson2014} extends OMP and succeeds to estimate \ofg parameters. 
COMP also operates with a parameter grid, 
but uses linear combinations of multiple atoms to approximate all atoms $\bs a(r,v)$ from the continuous dictionary. 
In other terms, it interpolates from the grid the atoms of $\cl D$ that are parameterized from \ofg parameters.
Our algorithm F-COMP applies the same interpolation concept to the atoms $\bs A(r, v)$, which are factorized by \eqref{eq_fact_atom}.

More precisely, let us define the separable grid $\OP = \{\omp_{n}\}_{n=1}^{N}$ $\subset \cl P$ such that $\OP = \OR \times \OV$, with $\OR := \{\omr_{n_\rr}\}_{n_\rr=1}^{N_\rr}\subset \cl R$ and $\OV:= \{\omv_{n_\rv}\}_{n_\rv=1}^{N_\rv}\subset \cl V$, respectively the range and speed grids. 
We have $N:=N_\rr N_\rv$ and the indices $n$ and $n_\rr, n_\rv$ are linked by $\omp_{n_\rv N_\rr + n_\rr} = (\omr_{n_\rr}, \omv_{n_\rv})^\top$.

In COMP, each atom $\bs a(r_k, v_k)$ in the model \eqref{eq_complete_y} is approximated from one of the grid bins with a generic interpolation model, inspired by~\cite{simoncelli2011, knudson2014}, that reads
\begin{equation}
    \ts \bs a(r_k, v_k) \simeq \sum_{i=1}^I c_k^{(i)}\bs d^{(i)}[n(k)],
    \label{eq_interp}
\end{equation}
where $\bs d^{(i)}[n(k)]$ is the $i$-th interpolant atom associated to the $n(k)$-th bin from the grid $\OP$, and $n(k)$ is a grid index which depends on the interpolation scheme and on $(r_k, v_k)$. 
The coefficients $c_k^{(i)}$ are obtained from a mapping function~\cite{knudson2014} denoted by $\cl C_{n}(r, v)$ and defined from the interpolation scheme.
More precisely, for all $k\in [K]$, $(c_k^{(1)}, \cdots c_k^{(I)}) = \cl C_{n(k)}(r_k, v_k)$.

In F-COMP, we aim to apply the same interpolation concept to the factorized model \eqref{eq_fact_atom}. To that end, we propose a ``factorization over interpolation" strategy where each $\bs A(r_k, v_k)$ is interpolated by
\begin{equation}
    \ts \bs A(r_k, v_k) \simeq \sum_{i=1}^I \tilde c_k^{(i)}\bs D^{(i)}[n_\rr(k), n_\rv(k)].
    \label{eq_interp_fact}
\end{equation}
In this scheme, for all $i\in[I]$, we decompose the global interpolant atoms $\bs D^{(i)}[n_\rr(k), n_\rv(k)]$ using interpolant sub-atoms denoted by $\bs\xi^{(i)}[n_\rr(k)]$ and $\bs\eta^{(i)}[n_\rv(k)]$, \ie
\begin{equation}
    \bs D^{(i)}[n_\rr(k), n_\rv(k)] = \bs\xi^{(i)}[n_\rr(k)] \big(\bs\eta^{(i)}[n_\rv(k)]\big)^\top.
    \label{eq_interp_fact2}
\end{equation}
As stated in the previous section, $\bs A(r, v)$ approximates the matrix-reshaped $\bs a(r,v)$. 
Therefore, the right-hand side of \eqref{eq_interp_fact} approximates the matrix-reshaped right-hand side of \eqref{eq_interp}.
\subsection{Order-1 Taylor approximation:} 
We restrict our study to the simple case where the interpolation is obtained from a Taylor approximation of the atoms of $\cl D$. We postpone to future work the application to radars of other interpolation schemes, such as polar interpolation~\cite{duarte2013, champagnat2019}. 
We approximate the atoms of $\cl D$ by an order-1 Taylor expansion which is described for all $n\in[N]$ from \eqref{eq_interp} by
\begin{equation}
    \ts \bs d^{(1)}[n] = \bs a(\omp_n), \nonumber
\end{equation}
\begin{equation}
    \ts \bs d^{(2)}[n] = \tilde{R}\frac{\partial \bs a}{\partial r} (\omp_n), \hspace{3mm} \bs d^{(3)}[n] = \tilde{V}\frac{\partial \bs a}{\partial v} (\omp_n).
    \label{eq_taylor1_exact}
\end{equation}
\begin{equation}
    \ts \cl C_{n_\rv N_\rr + n_\rr}(r, v) = \big(1,\hspace{1mm} \tilde{R}^{-1}(r-\omr_{n_\rr}) ,\hspace{1mm}\ts \tilde{V}^{-1}(v-\omv_{n_\rv}) \big).
    \label{eq_mapping_t1}
\end{equation}
where $\tilde{R}$ and $\tilde{V}$ are constants for dimensionality normalization purpose. Similarly, from the factorized model \eqref{eq_interp_fact} -  \eqref{eq_interp_fact2}, for all $(n_\rr, n_\rv)\in[N_\rr]\times [N_\rv]$, we set
\begin{align}
    \ts \bs \xi^{(1)}[n_\rr] & = \bs \xi^{(3)}[n_\rr] = \bs \psi(\omr_{n_\rr}), \nonumber \\
    \bs \eta^{(1)}[n_\rv] & = \bs \eta^{(2)}[n_\rv] = \bs \phi (\omv_{n_\rv}),  
    \label{eq_taylor1_fact}
\end{align}
\begin{equation}
    \ts \bs \xi^{(2)}[n_\rr] = \tilde{R}^{-1}\bpsi'(\omr_{n_\rr}), \hspace{4mm} \bs \eta^{(3)}[n_\rv] = \tilde{V}^{-1}\bphi'(\omr_{n_\rv}).\nonumber
\end{equation}
In this simple case, the interpolating coefficients in \eqref{eq_interp_fact} are identical to the coefficients in \eqref{eq_interp}, \ie  $\tilde c^{(i)}_k =  c^{(i)}_k$ for all $(k, i)\in[K]\times [I]$ and are obtained from \eqref{eq_mapping_t1}.

The models we just defined from the grids $\OR$ and $\OV$ enable us, in the next section, to derive algorithms that recover parameters $\{(r_k, v_k)\}_{k=1}^K$ lying off-the-grid.
This is done either with higher complexity from \eqref{eq_interp} or faster but less accurately from \eqref{eq_interp_fact}.


\begin{algorithm}[tb!]
\SetKwInOut{Input}{Input}\SetKwInOut{Output}{Output}
\SetKwFor{While}{While}{:}{end}

\Input{$K$, $\bs y$, $\big\{\bs d^{(i)}[n]\big\}_{(i,n)\in[I]\tm [N]}$, $\OP$.}

\Output{$\{\hat{\alpha}_k\}_{k=1}^K, \{(\hat{r}_k, \hat{v}_k)\}_{k=1}^K$} 

\Begin{

 Initialization: $\bs{r}^{(1)} = \bs{y}$, $k=1$;

\While{$k \leq K$}{
    
    \begin{flalign}
    & \hat{n}_k = \amin{n\in[N]}\big(\underset{\bs \beta\in\bb C^I}{\text{min}} \big\|\sum_{i=1}^I\beta_i \bs d^{(i)}[n]-\bs r^{(k)}\big\|_2^2 \big) &
    \label{eq_indsel_alg}
    \end{flalign}
    
    \begin{flalign}
    & \big\{\hat{\bs \beta}_{k'}\big\}_{k'=1}^k \hspace{-1mm} = \hspace{-3mm}\amin{\{\hat{\bs \beta}_{k'}\in \bb C^I\}_{k'=1}^k}\hspace{-2.5mm} \Big\|\hspace{-0.5mm} \sum_{k'=1}^k \sum_{i=1}^I \beta_{k'}^{(i)} \bs d^{(i)}\big[\hat{n}_{k'}\big] - \bs y \Big\|_2^2 & 
    \end{flalign}
    
    \begin{flalign}
    & \bs r^{(k+1)} = \bs y - \sum_{k'=1}^k \sum_{i=1}^I \hat \beta_{k'}^{(i)} \bs d^{(i)}\big[ \hat{n}_{k'}\big] &
    \end{flalign}
    
     $k \leftarrow k+1$
}
\vspace{2mm}
    \hspace{-2mm}for all $k\in[K]$,\\
    
    \begin{flalign}
        & (\hat{\alpha}_k, \hat{r}_k, \hat{v}_k) = \hspace{-1mm}\amin{\alpha\in\bb C, \bs (r,v)\in\cl P}\hspace{-2mm} \big\| \alpha \cl C_{\hat n(k)}(r, v) - \hat{\bs \beta}_k \big\|_2^2. &
        \label{eq_corr_step}
    \end{flalign}
}
  \caption{Continuous OMP (COMP) for Radar}
  \label{alg_COMP}
\end{algorithm}

\section{Off-the-grid Algorithms}
\label{sec_alg}

\subsection{Continuous OMP:} 


Alg. \ref{alg_COMP} formulates COMP for a generic interpolation scheme.
This algorithm adapts the methodology of OMP to greedily minimize $\big\|\bs y - \sum_{k=1}^K\alpha_k\sum_{i=1}^I c^{(i)}_k \bs d^{(i)}[n(k)]\big\|_2^2$.
The greedy iterations provide, for each target, the index $\hat n(k)$ of a corresponding on-the-grid estimator and a set of $I$ complex coefficients, instead of one in OMP. 
The recovered coefficients for the $k$-th target are gathered in $\hat{\bs \beta}_k := (\hat \beta^{(1)}_1, \cdots \hat \beta^{(I)}_k)$, where $\hat \beta_k^{(i)}$ estimates $\alpha_k c^{(i)}_k$.
These coefficients contain information on both $\alpha_k$ and the deviation of the target parameters from the grids. 
The correction step in \eqref{eq_corr_step} leverages this information from $\hat{\bs \beta}_k$ to compute \ofg estimators of $\{(r_k,v_k)\}_{k=1}^K$. 
Moreover, the joint computation of $\{\bs \beta_{k'}\}_{k'=1}^K$ in each iteration enables to adaptively adjust the effective \ofg values of the previous estimators.
This enables the distinction of closer targets with respect to OMP. 


The computational complexity of COMP is dominated by the index selection step. 
The inner minimization of \eqref{eq_indsel_alg} has a close form involving the product $(\bs d^{(1)}[n], \cdots, \bs d^{(i)})^H[n]\in  C^{I\times M}$ with $\bs r^{(k)}$.
Hence the complexity of \eqref{eq_indsel_alg} is $O(IMN)$.
Using the Taylor interpolation scheme we detailed in Sec. \ref{sec_sig} d), the index $n(k)$ is expected to correspond to the closest bin of $\OP$ to the parameter $(r_k, v_k)$. 
In that case, we can approximate the index selection of \eqref{eq_indsel_alg} by its corresponding formulation in OMP, whose complexity is $O(MN)$ and which is formulated by
\begin{equation}
    \ts \hat{n}_k = \amax{n\in[N]}|\sps{\bs d^{(1)}[n]}{\bs r^{(k)}}{}|.
    \label{eq_OMP_indsel}
\end{equation}
Advanced analysis of the implication of this simplification, as well as comparisons with the non-simplified implementation, are going to be presented in an extension of this contribution. 


\subsection{Factorized Continuous OMP:} 
Our algorithm, the Factorized Continuous OMP (F-COMP), leverages the factorized interpolated model \eqref{eq_interp_fact} to reduce the complexity of the estimation of the parameters. It follows the same steps as COMP and greedily minimizes $\big\|\bs Y - \sum_{k=1}^K\alpha_k\sum_{i=1}^I c^{(i)}_k \bs D^{(i)}[n_{\rr,k}]\big\|_F^2$.
We apply the same simplification of the index selection step \eqref{eq_OMP_indsel} that we used for COMP. From \eqref{eq_interp_fact2}, the indices $\hat n_\rr(k)$ and $\hat n_\rv(k)$ are selected by
\begin{equation}
    \ts (\hat n_\rr(k), \hat n_\rv(k)) = \hspace{-4mm}\amax{(n_\rr, n_\rv)\in[N_\rr]\times [N_\rv]}\hspace{-2mm} \big| (\bs\xi^{(1)}[n_\rr])^H \bs R^{(k)} (\bs \eta^{(1)}[n_\rv])^*\big|, \nonumber
    \label{eq_indsel_FOMP}
\end{equation}
where $\bs R^{(k)}$ is the residual of the $k$-th iteration. This maximization is computed with a complexity $O(N\min(M_c, M_s))$.


To leverage the factorization in the computation of $\{\hat{\bs \beta}_{k'}\}_{k'=1}^K$, we propose the following procedure, inspired by the implementation of the Factorized 2D-OMP (F-OMP) in~\cite{fang2015},
\begin{equation}
    (\hat{\bs \beta}_{1}^\top, \cdots, \hat{\bs \beta}_{k}^\top) = \bs H^{-1}\bs f,
\end{equation}
where $\bs H = \bs H_\xi\odot \bs H_\eta \in \bb C^{Ik\times Ik}$ where $\bs H_\xi = \bs \Xi^H\bs \Xi$ with
\begin{equation}
    \bs\Xi = \Big(\bs \xi^{(1)}[\hat n_{\rr,1}], \cdots,\bs\xi^{(I)}[\hat n_{\rr,1}],\cdots \bs\xi^{(I)}[\hat n_{\rr,k}]\Big)
\end{equation}
and $\bs f = (f_1^{(1)}, \cdots f_1^{(I)}, f_2^{(1)}, \cdots, f_k^{(I)})\in \bb C^{Ik}$ where
\begin{equation}
    f_{k'}^{(i)} = (\bs\xi^{(i)}[\hat n_{\rr,k'}])^H\bs Y (\bs\eta^{(i)}[\hat n_{\rv,k'}])^*.
\end{equation}

The implementations above enable F-COMP to estimate the target parameters faster than COMP. 
Still, this estimation is expected to be less accurate than COMP because the factorized model in \eqref{eq_Yfact} --- from which F-COMP derives --- is an approximation of the exact radar signal \eqref{eq_complete_y}. 


\newlength{\hlabel}
\setlength{\hlabel}{4.82cm}
\begin{figure*}[tb]
\begin{minipage}{0.68\linewidth}
    \includegraphics[width=\textwidth]{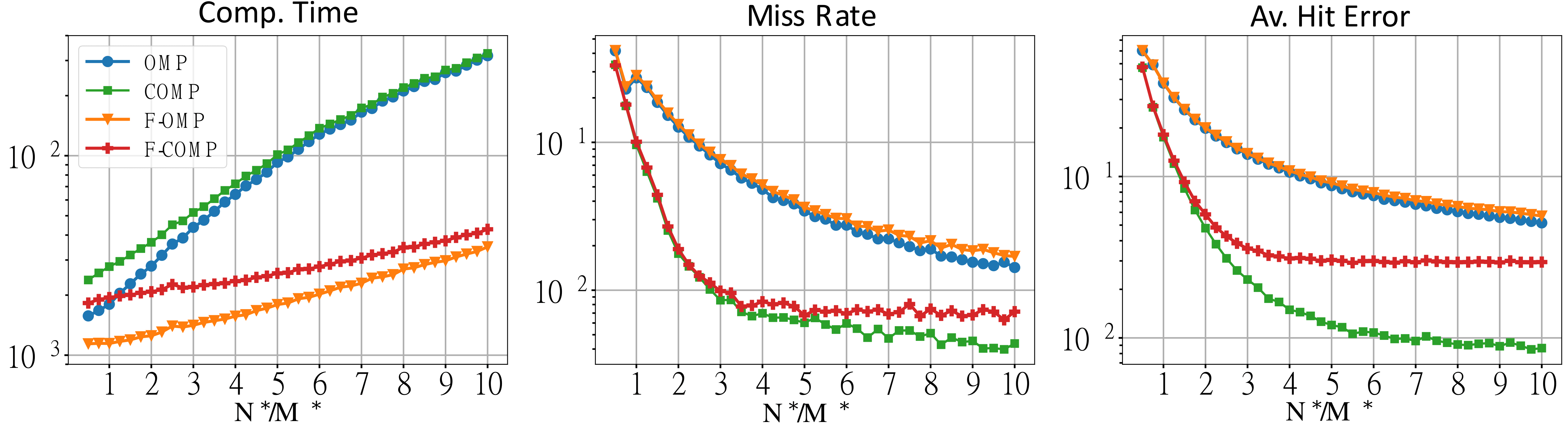}
    \caption{Comparison of (a) Computation Time (seconds), (b) Miss Rate and (c) Average Error Within Successes of OMP, COMP, F-OMP and F-COMP in function of the number of bins the location and in the velocity grids ($N^* = N_\rr = N_\rv$). The simulated system has $M^*=16$.}
    \label{fig_comp4algo}
    \vspace{-\hlabel}
    \vspace{\hlabel}
    \includegraphics[width=\textwidth]{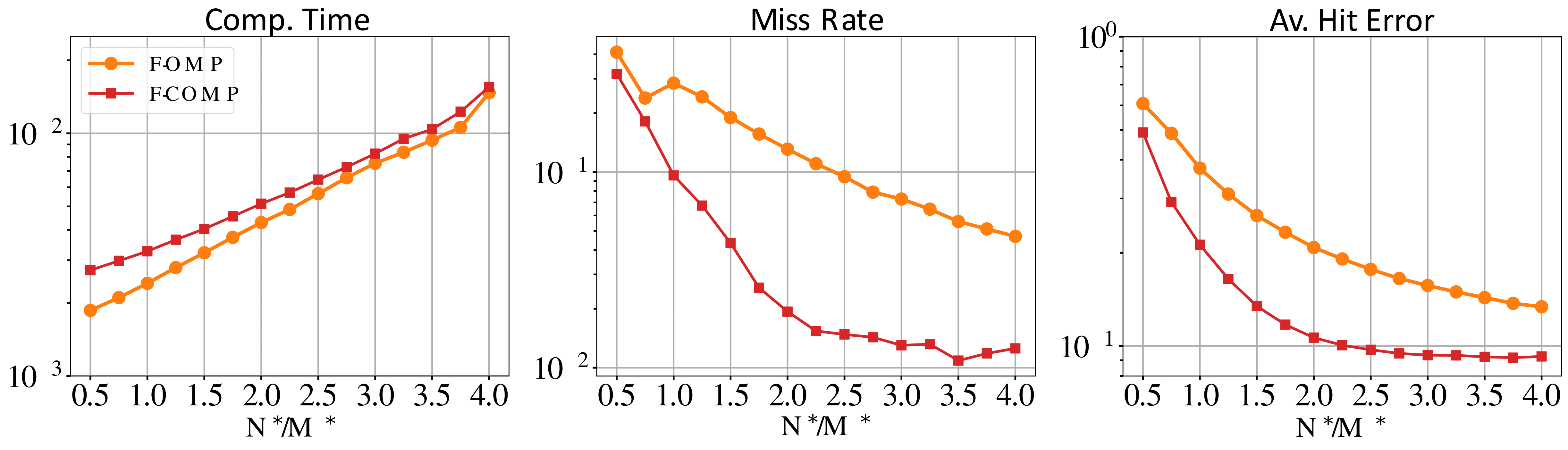}
    \caption{Comparison of (a) Computation Time, (b) Miss Rate and (c) AHE of F-OMP and F-COMP in function of the number of bins in the location and velocity grids and with $M^*=64$.}
    \vspace{-4.8cm}
    \vspace{5cm}
    \label{fig_comp2algo_fft}
    \vspace{-0mm}
\end{minipage}
\hspace{2mm}
\begin{minipage}{0.3\linewidth}
    \vspace{-10cm}
    \hspace{8mm}\includegraphics[width=0.62\textwidth]{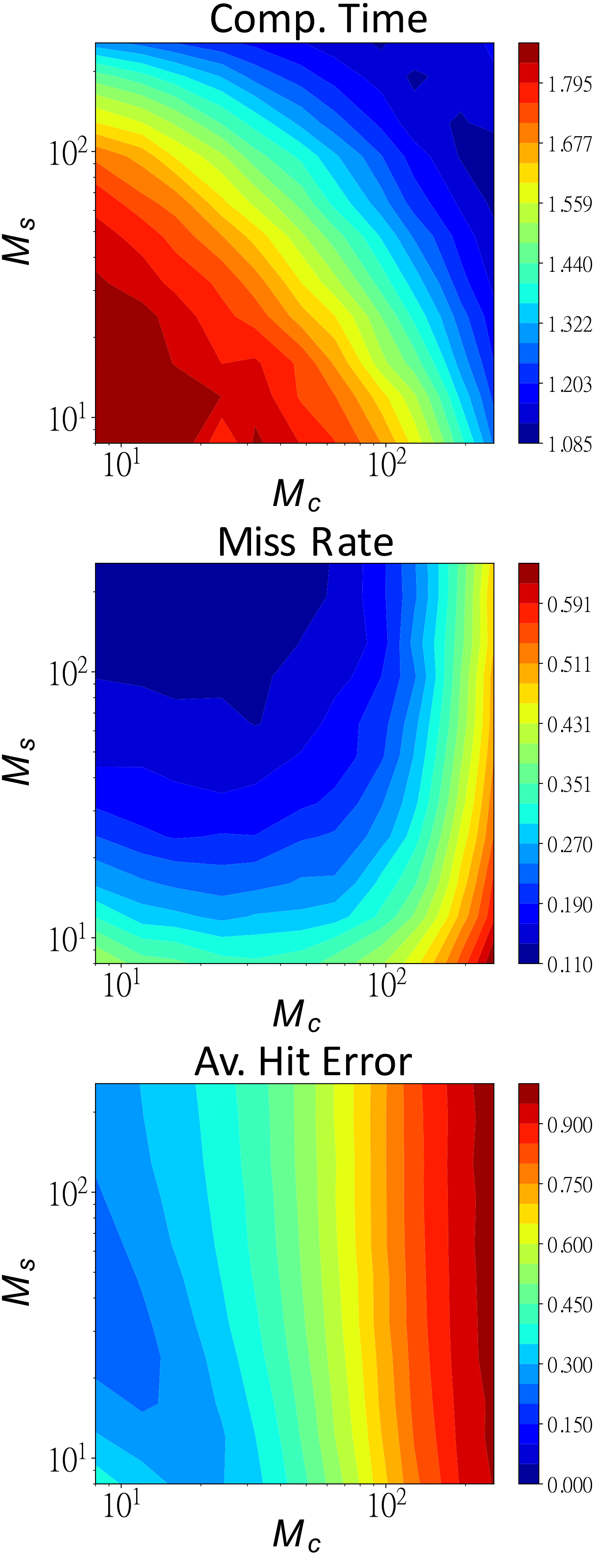}
    \caption{(a) $\frac{\text{Comp. Time of F-COMP}}{\text{Comp. Time of F-OMP}}$, (b) $\frac{\text{MR of F-COMP}}{\text{MR of F-OMP}}$ and (c) $\frac{\text{AHE of F-COMP}}{\text{AHE of F-OMP}}$ in function of $M_s$ and $M_c$.}
    \label{fig_paramtest}
    \vspace{-10cm}
\end{minipage}
\end{figure*}

\subsection{Off-the-grid correction:}
The final step of both F-COMP and COMP is the correction step formulated by \eqref{eq_corr_step}. For a given $k\in[K]$, we estimate the scattering coefficient $\alpha_k$ and the \ofg deviations that we denote by $\delta_\rr := \tilde{R}^{-1}(r_k-\omr_{\hat n_\rr(k)})$ and $\delta_\rv := \tilde{V}^{-1}(v_k-\omv_{\hat n_\rv(k)})$.
We implement this step for the order-1 Taylor interpolation scheme we described in Sec. \ref{sec_sig}.
The substitution of \eqref{eq_mapping_t1} in \eqref{eq_corr_step} leads to the following coupled relations,
\begin{equation}
   \ts \hat \alpha_k = \frac{\hat \beta_k^{(1)} + \hat \beta_k^{(2)}\hat \delta_\rr + \hat \beta_k^{(3)}\hat \delta_\rv}{1 + \hat \delta_\rr^2 + \hat \delta_\rv^2}. 
    \label{eq_c2alpha}
\end{equation}
\begin{equation}
    \ts \hat \delta_\rr = \Re\{\hat \beta_k^{(2)} / \hat \alpha_k\}, \hspace{2mm} \hat\delta_\rv = \Re\{\hat \beta_k^{(3)} / \hat \alpha_k\}. 
    \label{eq_c2delta}
\end{equation}
The coupled equations above are easy to solve by iteratively computing \eqref{eq_c2alpha} and \eqref{eq_c2delta} with the intialization $\hat \delta_\rr = \hat \delta_\rv = 0$.
Then, for each $k\in[K]$, we directly compute $\hat r_k$ and $\hat v_k$ from
\begin{equation}
    \ts (\hat r_k, \hat v_k)^\top = \omp_{\hat n(k)} + \big( \tilde{R} \hat \delta_\rr, \tilde{V}\hat \delta_\rv  \big)^\top, 
\end{equation}
which ends the algorithms COMP and F-COMP.

\section{Numerical Results}
\label{sec_sim}

\subsection{Simulated system and performance metrics:}
The effectiveness of F-COMP with respect to both COMP and F-OMP is validated with simulated signals from an FMCW radar as defined in Sec. \ref{sec_sig}.
The simulated system is characterized by $B = 200$MHz, $f_0=24$GHz, $T_s=5\mu$s and $T_c = M_sT_s$. 

We study the evolution of the computation time and estimation performance with respect to the number of grid bins $N^* := N_\rr = N_\rv$. 
The performance is evaluated with (i) the Miss Rate (MR) and (ii) the Average Hit Error (AHE).
They are computed from the estimation error which, for each $k\in[K]$ in a given realisation of $K$ targets \footnote{For a given realisation of $K$ targets, the estimators $\{\hat r_k, \hat v_k\}_{k=1}^K$ are sorted and associated to the exact values $\{ r_k, v_k\}_{k=1}^K$ without repetition and such that the miss rate is the smallest. }, is defined by 
\begin{equation}
    E_k = \ts \sqrt{\big(\frac{\hat r_k - r_k}{\rho_\rr}\big)^2 + \big(\frac{\hat v_k - v_k}{\rho_\rv}\big)^2}.
    \label{eq_esterror}
\end{equation}
In \eqref{eq_esterror}, $\rho_\rr := \frac{\cel}{2B}$ and $\rho_\rv := \frac{\cel}{4f_0M_cT_c}$ characterize the resolutions of, respectively, the range and the speed estimations.
More precisely, they correspond to the width of the main lobe of the cardinal sine-shaped approximated ambiguity functions. 
The $k$-th estimator is a \emph{miss} if $E_k>1$.
Each dot of all curves is obtained by averaging the values of the metrics resulting from the application of the different algorithms on 10,000 realisations of random sets of $K=5$ independent targets.
Those realisations are characterized by $\alpha_k \sim \bb C\cl N(0, 1)$ and $r_k$ (resp. $v_k$) uniformly chosen in $\cl R$ (resp. $\cl V$) for all $k\in[K]$. 
We set $\cl R = ]0, M_s\frac{c}{2B}]$ (meters) and $\cl V = ]-\frac{\cel}{4f_0T_c}, \frac{\cel}{4f_0T_c}]$ (meters per second).

\subsection{Comparative results:}
In real radar applications, the ``non-factorized algorithms" (OMP and COMP) which exploit the exact signal model are often not practicable because of their excessively high memory and time requirements.
Before comparing the two ``factorized algorithms" (F-OMP and F-COMP) with realistic radar parameter's values,  we first compared the results of all four algorithms (F-)(C)OMP with low numbers of acquired samples, \ie $M^* := M_s=M_c=16$. 
This enables separable observations of the effect of the factorization and the interpolation. 

From Fig. \ref{fig_comp4algo}(a), it is clear that the computation times of non-factorized algorithms grow faster with $N^*$ than the factorized ones. 
The gaps in computation time between the \emph{Continuous} algorithms (F-COMP and COMP) and their respective \ong counterpart (F-OMP and OMP) correspond the computation time of the \ofg corrections. 
Fig. \ref{fig_comp4algo}(b) and (c) show that the continuous algorithms provide better estimations than \ong algorithms for all values of $N^*$. 

The performance loss caused by the approximation used by the factorized algorithms only appears with a dense grid (high value of $N^*/M^*$).
Indeed, Fig. \ref{fig_comp4algo}(b) and (c) shows that when the grid density is low, F-COMP and COMP almost have identical performance because the effect of this approximation is dominated by the grid error.
Therefore, non-factorized algorithms are only advantageous when using a dense grid, in which case they are not practicable in term of computation time and memory requirement.
Regarding the factorized algorithms, F-COMP is only slower than F-OMP by a constant factor (which depends of the interpolation scheme) while providing significant improvement by enabling off-the-grid target estimation.
For this reason F-COMP appears as the best trade-off between performance and computation time for most values of MR and AHE it can reach.

Fig. \ref{fig_comp2algo_fft} presents the result from a simulated radar with the more realistic value $M^*=64$. 
Fig. \ref{fig_comp2algo_fft} (b) and (c) reveal that the F-COMP curves saturate to greater (\ie worse) values of MR and AHE than the results in Fig. \ref{fig_comp4algo}.
This deterioration is caused by an increase of the model mismatch because $M_c\frac{B}{f_0}$ is larger, which diminishes the validity of neglecting the distorsion in \eqref{eq_Yfact}. 


\subsection{Impact of system parameters:}
We analysed the effect of $M_s$ and $M_c$ on the performance gain between F-COMP and F-OMP.
This effect is due to their impact on the model mismatch. 
We varied these numbers from 8 to 256 while maintaining $N_\rr = 2M_s$ and $N_\rv = 2M_c$. 
Fig. \ref{fig_paramtest} (a), (b) and (c) show the ratios respectively between, the computation times, the MR and the AHE of F-COMP and F-OMP.
Increasing $M_s$ (resp. $M_c$) improves the resolution of the range (resp. speed) estimation and leads to a reduction of the miss rate of F-COMP with respect to F-OMP (Fig. \ref{fig_paramtest} (b)). 
Yet, this improvement is dominated by the increasing model mismatch as $M_c$ gets larger.
This causes the deterioration of the MR and AHE ratios in Fig \ref{fig_paramtest} (b) and (c) for large values of $M_c$.
To sum up, Fig. \ref{fig_paramtest} reveals that for all values $M_s$ and $M_c$, F-COMP always exhibits better performance than F-OMP. Yet, the improvement tends to fade as $M_c \frac{B}{f_0}$ increases.

\section{Conclusion and Future Work}
In this contribution, we studied the detection of multiple \ofg moving targets using FMCW radars.
This motivated us to extend the Continuous OMP to factorized models, and hence to design the F-COMP. 
We showed that F-COMP gathers the advantages of both COMP and F-OMP and provides the best trade-off between complexity and accuracy of estimation. 
In radar applications, we showed how some values of the radar system parameters affect, but do not remove, the performance gain of F-COMP over F-OMP. 
This paper proposed a simple implementation of the concepts of F-COMP. 
Several enhancements can increase its effectiveness over other greedy algorithms. 
More precisely, in future work we may investigate the extension of more sophisticated interpolation schemes to factorizable dictionaries, and study the compromise between the interpolation order and the required grid density for optimal correction.
Moreover, the presented algorithms can be extended to include mismatch compensation~\cite{feuillen2016} or adaptive grid~\cite{liu2018} strategies.

\appendix
\bibliographystyle{unsrt}
\begin{small}
\bibliography{biblio.bib}
\end{small}

\end{document}